\documentclass[12pt]{article}
\usepackage{pdproc,epsfig} 

\def\dd{\displaystyle}

\def\bea{\begin{eqnarray}}
\def\eea{\end{eqnarray}}
\def\beq{\begin{equation}}
\def\eeq{\end{equation}}
\def\bq{\begin{quote}}
\def\eq{\end{quote}}
\def\gappeq{\mathrel{\rlap {\raise.5ex\hbox{$>$}} {\lower.5ex\hbox{$\sim$}}}}
\def\lappeq{\mathrel{\rlap{\raise.5ex\hbox{$<$}} {\lower.5ex\hbox{$\sim$}}}}

\def\GeV{{\rm GeV}}
\def\be{\begin{equation}}
\def\ee{\end{equation}}
\def\bc{\begin{center}}
\def\ec{\end{center}}
\def\bea{\begin{eqnarray}}
\def\eea{\end{eqnarray}}
\def\dd{\displaystyle}

\def\gappeq{\mathrel{\rlap {\raise.5ex\hbox{$>$}} {\lower.5ex\hbox{$\sim$}}}}
\def\lappeq{\mathrel{\rlap{\raise.5ex\hbox{$<$}} {\lower.5ex\hbox{$\sim$}}}}
\newcommand{\bac}{\beq\begin{array}}
\newcommand{\eac}{\end{array}\eeq}
\newcommand{\ba}{\begin{array}}
\newcommand{\ea}{\end{array}}
\newcommand{\beaa}{\begin{eqnarray*}}
\newcommand{\eeaa}{\end{eqnarray*}}

\begin{document}

\vspace*{-1cm}
\phantom{hep-ph/***}

\hfill{RM3-TH/11-17~~~~~~~~~~~~~~~~~~~~~~~~~~~~ CERN-PH-TH/2011-298~~~~~~~~~~~~~~~~}

\vskip 2.5cm

\renewcommand{\thefootnote}{\alph{footnote}}
  
\title{The Mystery of Neutrino Mixings}

\author{Guido Altarelli}

\address{Dipartimento di Fisica, Universita' di Roma Tre\\
Rome, Italy\\
and\\
CERN, Department of Physics, Theory Division \\  
CH-1211 Gen\`eve 23, Switzerland\\
{\rm E-mail: guido.altarelli@cern.ch}}

\abstract{In the last years we have learnt a lot about neutrino masses and mixings.  Neutrinos are not all massless but their masses are very small. Probably masses are small because neutrinos are Majorana particles
with masses inversely proportional to the large scale M of lepton number (L) violation, which turns out to be compatible with the GUT scale. We have understood that there is no contradiction between large neutrino mixings and small quark mixings, even in the context of GUTs and that neutrino masses fit well in the SUSY GUT picture. Out of equilibrium decays with CP and L violation of heavy RH neutrinos can produce a B-L asymmetry, then converted near the weak scale by instantons into an amount of B asymmetry compatible with observations (baryogenesis via leptogenesis).  It appears that active neutrinos are not a significant component of Dark Matter in the Universe.   A long list of models have been formulated over the years to understand neutrino masses and mixings. With the continuous improvement of the data most of the models have been discarded by experiment. The surviving models still span a wide range going from a maximum of symmetry, with discrete non-abelian flavour groups, to the opposite extreme of anarchy.}

\normalsize\baselineskip=15pt

\vskip 2cm

\section{Introduction}

Experiments on neutrino oscillations, which measure differences of squared masses as well as mixing angles \cite{review} \cite{rev2} have established that neutrinos have a mass. Two distinct oscillation frequencies have been first measured in solar and atmospheric neutrino oscillations and later confirmed by experiments on earth, like KamLAND, K2K and MINOS. Two well separated differences need at least three different neutrino mass eigenstates involved in oscillations so that the three known neutrino species can be sufficient. Then at least two $\nu$'s must be massive while, in principle, the third one could still be massless. A signal corresponding to a third mass difference was claimed by the LSND experiment (with antineutrinos) but not confirmed by KARMEN. More recently MiniBooNE \cite{Mini} has reported some possible supporting evidence for the LSND effect in their antineutrino run while no oscillation is observed in the neutrino run.  The existence of a third oscillation frequency would imply the need for additional sterile neutrinos (i.e. with no weak interactions, as any new light active neutrino was excluded by LEP) or CPT violation (as, in this case, the masses of neutrinos and antineutrinos can be different). 

The main recent developments on the experimental side were the results on $\theta_{13}$ from T2K \cite{t2k} and MINOS \cite{min} (very recently also DOUBLE CHOOZ \cite{DC}) and the coming back of sterile neutrinos. As well known, the T2K run was suddenly interrupted by the devastating earthquake that hit Japan on March 11, just minutes away from the scheduled presentation of the first T2K data. Later T2K released the first publication on their data  \cite{t2k}, reporting a 2.5$\sigma$ signal  for $\sin^2{2\theta_{13}}$ that indicates a value of $\theta_{13}$ close to the previous upper bound, of the order of the Cabibbo angle $\theta_C$.

On the evidence for sterile neutrinos a number of hints have been recently reported. They do not make yet a clear evidence but certainly pose an experimental problem that needs clarification. First, there is the MiniBooNE experiment \cite{min} that in the antineutrino channel reports an excess of events supporting the LSND oscillation signal (originally observed with antineutrinos). More recently an update of the MiniBooNE data in the antineutrino channel shows less supporting evidence \cite{Mini11}.   In the neutrino channel MiniBooNE did not observe a signal in the LSND domain. However, in these data there is a unexplained excess at low energy over the (reliably?) estimated background.  In the neutrino data sample, for the search of a LSND-like signal, only the events with neutrino energy above a threshold value $E_{th}$ were used, leaving the issue of an explanation of the low energy excess unanswered. In the antineutrino channel most of the support to the LSND signal appears to arise from an excess above $E_{th}$ but quite close to it, so that there is, in my opinion, some room for perplexity. Then there is the reactor anomaly: a reevaluation of the reactor flux \cite{flux} produced an apparent gap between the theoretical expectations and the data taken at small distances from the reactor ($\lappeq$ 100 m). The discrepancy is of the same order of the quoted systematic error whose estimate, detailed in the paper, should perhaps be reconsidered. Similarly the Gallium anomaly \cite{gal} depends on the assumed cross-section which could be questioned. The reactor anomaly and the Gallium anomaly do not really agree on the oscillation parameters that they point to: the $\Delta m^2$ values are compatible but the central values of $\sin^2{2\theta}$ differ by about an order of magnitude, with Gallium favouring the larger angle. Cosmological data allow the existence of one sterile neutrino, while the most stringent bounds arising form nucleosynthesis disfavour two or more sterile neutrinos \cite{cosmo}. Over all, only a small leakage from active to sterile neutrinos is allowed by present neutrino oscillation data, as discussed in refs. \cite{ste} If all the indications listed above were confirmed (it looks unlikely) then one sterile neutrino would not be enough and at least two would be needed with sub-eV masses. Establishing the existence of sterile neutrinos would be a great discovery. In fact a sterile neutrino is an exotic particle not predicted by the most popular models of new physics. A sterile neutrino is not a 4th generation neutrino: the latter is coupled to the weak interactions (it is active) and heavier than half the Z mass. A sterile neutrino would probably be a remnant of some hidden sector. The issue is very important so that new and better experimental data are badly needed. 

As already mentioned, in neutrino oscillations the leakage from the three active species towards the sterile neutrinos is any case small and, in fact, the best established oscillation phenomena are well described in terms of 3-neutrino models. In this domain the main recent developments have been the T2K and MINOS results on $\theta_{13}$.   The T2K result \cite{t2k}, based on the observation of 6 electron events when $1.5\pm0.3$ are expected for $\theta_{13}=0$, is converted into a confidence interval $0.03(0.04)\le \sin^2{2\theta_{13}} \le 0.28(0.34)$ at $90\%$ c.l. for $\sin^2{2\theta_{23}}=1$, $|\Delta m^2| = 2.4~10^{-3} eV^2$, $\delta_{CP} = 0$ and for normal (inverted) neutrino mass hierarchy. Also the MINOS Collaboration released \cite{min} their corresponding $90\%$ c.l. range
as $0(0)\le \sin^2{2\theta_{13}} \le 0.12(0.19)$, which is displaced towards smaller values with respect to that of T2K. Finally DOUBLE CHOOZ \cite{DC} finds (with only the far detector in operation): $\sin^2{2\theta_{13}}=0.085\pm0.051$ at 1$\sigma$.

\section{Neutrino Masses and Lepton Number Violation}

Neutrino oscillations imply non vanishing neutrino masses which in turn demand either the existence of right-handed (RH) neutrinos
(Dirac masses) or lepton number L violation (Majorana masses) or both. Given that neutrino masses are certainly extremely
small, it is really difficult from the theory point of view to avoid the conclusion that L conservation must be violated.
In fact, in terms of lepton number violation the smallness of neutrino masses can be explained as inversely proportional
to the very large scale where L is violated, of order $M_{GUT}$ or even $M_{Pl}$.

If L conservation is violated neutrinos can be Majorana fermions. For a Majorana neutrino each mass eigenstate with given helicity coincides with its own antiparticle with the same helicity. As well known, for a charged massive fermion there are four states differing by their charge and helicity (the four components of a Dirac spinor) as required by Lorentz and CPT invariance. For a massive Majorana neutrino, neutrinos and antineutrinos can be identified and only two components are needed to satisfy the Lorentz and CPT invariance constraints. Neutrinos can be Majorana fermions because, among the fundamental fermions (i.e. quarks and leptons),  they are the only electrically neutral ones. If, and only if, the lepton number L is not conserved, i.e. it is not a good quantum number, then neutrinos and antineutrinos can be identified. For Majorana neutrinos both Dirac mass terms, that conserve L ($\nu \rightarrow \nu$), and Majorana mass terms, that violate L by two units ($\nu \rightarrow \bar{\nu}$), are in principle possible. Of course the restrictions from gauge invariance must be respected. So for neutrinos the Dirac mass terms ($\bar{\nu}_R\nu_L$ +h.c.) arise from the couplings with the Higgs  field, as for all quarks and leptons. For Majorana masses, a $\nu_L^T \nu_L$ mass term has weak isospin 1 and needs two Higgs fields to make an invariant. On the contrary  a $\nu_R^T \nu_R$ mass term is a gauge singlet and needs no Higgs. As a consequence, the right-handed neutrino Majorana mass $M_R$ is not bound to be of the order of the electroweak symmetry breaking (induced by the Higgs vacuum expectation value) and can be very large (see below).

Some notation: the charge conjugated of $\nu$ is $\nu^c$, given by $\nu^c = C(\bar{\nu})^T$, where $C=i\gamma_2 \gamma_0$ is the charge conjugation matrix acting on the spinor indices (in the following, when dealing with the flavour structure of couplings, the $C$ matrix will be often omitted but understood). In particular $(\nu^c)_L = C(\bar{\nu_R})^T$, so that, instead of using $\nu_L$ and $\nu_R$, we can refer to $\nu_L$ and $(\nu^c)_L $, or simply $\nu$ and $\nu^c$. 

Once we accept L non-conservation we gain an elegant explanation for the smallness of neutrino masses. If L is not
conserved, even in the absence of heavy RH neutrinos, Majorana masses for neutrinos can be generated by dimension five
operators \cite{weinberg} of the form 
\beq 
O_5=\frac{(H l)^T_i \lambda_{ij} (H l)_j}{\Lambda}~~~,
\label{O5}
\eeq  
with $H$ being the ordinary Higgs doublet, $l_i$ the SU(2) lepton doublets, $\lambda$ a matrix in  flavour space,
$\Lambda$ a large scale of mass, of order $M_{GUT}$ or $M_{Pl}$ and a charge conjugation matrix $C$
between the lepton fields is understood. 
Neutrino masses generated by $O_5$ are of the order
$m_{\nu}\approx v^2/\Lambda$ for $\lambda_{ij}\approx {\rm O}(1)$, where $v\sim {\rm O}(100~\GeV)$ is the vacuum
expectation value of the ordinary Higgs.

We consider that the existence of RH neutrinos $\nu^c$ is quite plausible because most GUT groups larger than SU(5) require
them. In particular the fact that $\nu^c$ completes the representation 16 of SO(10): 16=$\bar 5$+10+1, so that all
fermions of each family are contained in a single representation of the unifying group, is too impressive not to be
significant. At least as a classification group SO(10) must be of some relevance in a more fundamental layer of the theory! Thus in the following we both assume that
$\nu^c$ exist and L is not conserved. With these assumptions the see-saw mechanism \cite{seesaw} is possible.   We recall, also to fix notations, that in its simplest form it arises as follows. Consider the SU(3) $\times$ SU(2) $\times$ U(1)
invariant Lagrangian giving rise to Dirac and $\nu^c$ Majorana masses (for the time being we consider the $\nu$
(versus $\nu^c$) Majorana  mass terms as comparatively negligible):
\beq 
{\cal L}=-{\nu^c}^T y_\nu (H l)+\frac{1}{2}
{\nu^c}^T M \nu^c +~h.c.
\label{lag}
\eeq  
The Dirac mass matrix $m_D\equiv y_\nu v/\sqrt{2}$, originating from electroweak symmetry breaking,  is, in general,
non-hermitian and non-symmetric, while the Majorana mass matrix $M$ is symmetric,
$M=M^T$. We expect the eigenvalues of $M$ to be of order $M_{GUT}$ or more because $\nu^c$ Majorana masses are
SU(3)$\times$ SU(2)$\times$ U(1) invariant, hence unprotected and naturally of the order of the cutoff of the low-energy
theory.  Since all $\nu^c$ are very heavy we can integrate them away.  For this purpose we write down the equations of
motion for $\nu^c$ in the static limit, $i.e.$ neglecting their kinetic terms:
\beq  -\frac{\partial {\cal L}}{\partial\nu^c}=y_\nu (H l)- M \nu^c= 0~~~.
\label{eulag}
\eeq  {}From this, by solving for $\nu^c$, we obtain:
\beq
\nu^c= M^{-1} y_\nu (H l)~~~.
\label{R}
\eeq  We now replace in the lagrangian, eq. (\ref{lag}), this expression for $\nu^c$ and we get the operator $O_5$ of eq.
(\ref{O5}) with
\beq 
\frac{2 \lambda}{\Lambda}=-y_\nu^T M^{-1} y_\nu ~~~~~,
\eeq and the resulting neutrino mass matrix reads:
\beq  m_{\nu}=m_D^T M^{-1}m_D~~~.
\eeq  This is the well known see-saw mechanism result \cite{seesaw}: the light neutrino masses are quadratic in the Dirac
masses and inversely proportional to the large Majorana mass.  If some $\nu^c$ are massless or light they would not be
integrated away but simply added to the light neutrinos. Notice that the above results hold true for any number
$n$ of heavy neutral fermions 
$R$ coupled to the 3 known neutrinos. In this more general case $M$ is an $n$ by $n$ symmetric matrix and the coupling
between heavy and light fields is described by the rectangular $n$ by 3 matrix $m_D$.  Note that for
$m_{\nu}\approx \sqrt{\Delta m^2_{atm}}\approx 0.05$ eV (see Table(\ref{tab:data})) and 
$m_{\nu}\approx m_D^2/M$ with $m_D\approx v
\approx 200~GeV$ we find $M\approx 10^{15}~GeV$ which indeed is an impressive indication for
$M_{GUT}$.

If additional non-renormalizable contributions to $O_5$, eq. (\ref{O5}), are comparatively non-negligible, they should
simply be added.  For instance in SO(10) or in left-right extensions of the SM, an SU(2)$_L$ triplet can couple to two 
lepton doublets and to two Higgs and may induce a sizeable contribution to neutrino masses. At the level of the 
low-energy effective theory, such contribution is still described by the operator $O_5$ of eq. (\ref{O5}),
obtained by integrating out the heavy SU(2)$_L$ triplet. This contribution is called type II
to be distinguished from that obtained by the exchange of RH neutrinos (type I). One can also have the exchange of a fermionic SU(2)$_L$ triplet coupled to a lepton doublet and a Higgs (type III).
After elimination of the heavy fields, at the level of the effective low-energy theory, the
three types of see-saw terms are equivalent. In particular they have identical transformation properties under a chiral change of
basis in flavour space. The difference is, however, that in type I see-saw mechanism, the Dirac matrix
$m_D$ is presumably related to ordinary fermion masses because they are both generated by the Higgs mechanism and both
must obey GUT-induced constraints. Thus more constraints are implied if one assumes the see-saw mechanism in its simplest type I version.
.

\section{Basic Formulae and Data for Three-Neutrino Mixing}

We assume in the following that there are only two distinct
neutrino oscillation frequencies, the atmospheric and the solar frequencies. These two can be reproduced with the known
three light neutrino species (for more than three neutrinos see, for example, ref.\cite{ste}). 

Neutrino oscillations are due to a misalignment between the flavour basis, $\nu'\equiv(\nu_e,\nu_{\mu},\nu_{\tau})$, where
$\nu_e$ is the partner of the mass and flavour eigenstate $e^-$ in a left-handed (LH) weak isospin SU(2) doublet (similarly
for 
$\nu_{\mu}$ and $\nu_{\tau}$) and the mass eigenstates $\nu\equiv(\nu_1, \nu_2,\nu_3)$ \cite{pon,lee}: 
\beq
\nu' =U \nu~~~,
\label{U}
\eeq  where $U$ is the unitary 3 by 3 mixing matrix. Given the definition of $U$ and the transformation properties of the
effective light neutrino mass matrix $m_{\nu}$ in eq. (\ref{O5}):
\bea 
\label{tr} {\nu'}^T m_{\nu} \nu'&= &\nu^T U^T m_\nu U \nu\\ \nonumber  U^T m_{\nu} U& = &{\rm
Diag}\left(m_1,m_2,m_3\right)\equiv m_{diag}~~~,
\eea  we obtain the general form of $m_{\nu}$ (i.e. of the light $\nu$ mass matrix in the basis where the charged lepton
mass is a diagonal matrix):
\beq  m_{\nu}=U^* m_{diag} U^\dagger~~~.
\label{gen}
\eeq  The matrix $U$ can be parameterized in terms of three mixing angles $\theta_{12}$,
$\theta_{23}$ and $\theta_{13}$ ($0\le\theta_{ij}\le \pi/2$)  and one phase $\varphi$ ($0\le\varphi\le 2\pi$) \cite{cab},
exactly as for the quark mixing matrix $V_{CKM}$. The following definition of mixing angles can be adopted:
\beq  U~=~ 
\left(\matrix{1&0&0 \cr 0&c_{23}&s_{23}\cr0&-s_{23}&c_{23}     } 
\right)
\left(\matrix{c_{13}&0&s_{13}e^{i\varphi} \cr 0&1&0\cr -s_{13}e^{-i\varphi}&0&c_{13}     } 
\right)
\left(\matrix{c_{12}&s_{12}&0 \cr -s_{12}&c_{12}&0\cr 0&0&1     } 
\right)
\label{ufi}
\eeq  where $s_{ij}\equiv \sin\theta_{ij}$, $c_{ij}\equiv \cos\theta_{ij}$.  In addition, if $\nu$ are Majorana particles,
we have the relative phases among the Majorana masses
$m_1$, $m_2$ and $m_3$.  If we choose $m_3$ real and positive, these phases are carried by $m_{1,2}\equiv\vert m_{1,2}
\vert e^{i\phi_{1,2}}$
\cite{frsm}.  Thus, in general, 9 parameters are added to the SM when non-vanishing neutrino masses are included: 3
eigenvalues, 3 mixing angles and 3 CP  violating phases.

In our notation the two frequencies, $\Delta m^2_{I}/4E$ $(I$=sun,atm), are parametrized in terms of the $\nu$ mass
eigenvalues by 
\beq
\Delta m^2_{sun}\equiv \vert\Delta m^2_{12}\vert ,~~~~~~~
\Delta m^2_{atm}\equiv \vert\Delta m^2_{23}\vert~~~.
\label{fre}
\eeq   where $\Delta m^2_{12}=\vert m_2\vert^2-\vert m_1\vert^2 > 0$ and $\Delta m^2_{23}= m_3^2-\vert m_2\vert ^2$. The
numbering 1,2,3 corresponds to our definition of the frequencies and in principle may not coincide with the ordering from
the lightest to the heaviest state. In fact, the sign of $\Delta m^2_{23}$ is not known and its determination is one of the existing experimental challenges. A positive (negative) sign corresponds to normal (inverse) hierarchy.

With the above definitions the present data are summarised in Table(\ref{tab:data}) \cite{Fogli}, \cite{Schwetz}.

\begin{table}[h]
\begin{center}
\begin{tabular}{|c|c|c|}
  \hline
  Quantity & Fogli et al \cite{Fogli} & Schwetz et al \cite{Schwetz} \\
  \hline
  $\Delta m^2_{sun}~(10^{-5}~{\rm eV}^2)$ &$7.58^{+0.22}_{-0.26}$ & $7.59^{+0.20}_{-0.18}$  \\
  $\Delta m^2_{atm}~(10^{-3}~{\rm eV}^2)$ &$2.35^{+0.12}_{-0.09}$ & $2.50^{+0.09}_{-0.16}$  \\
  $\sin^2\theta_{12}$ &$0.312^{+0.017}_{-0.016}$ & $0.312^{+0.017}_{-0.015}$ \\
  $\sin^2\theta_{23}$ &$0.42^{+0.08}_{-0.03}$ &  $0.52^{+0.06}_{-0.07}$ \\
  $\sin^2\theta_{13}$ &$0.025\pm0.007$ &$0.013^{+0.007}_{-0.005}$  \\
  \hline
  \end{tabular}
\end{center}
\begin{center}
\begin{minipage}[t]{12cm}
\caption{\label{tab:data}} Fits to neutrino oscillation data. The results correspond to the new reactor fluxes. The fit of Schwetz et al \cite{Schwetz} refers to the normal hierarchy case 
(in the inverse hierarchy case the main difference is that $\sin^2{\theta_{13}}= 0.016+0.008-0.006$)
\end{minipage}
\end{center}
\end{table}

Oscillation experiments do not provide information about the absolute neutrino mass scale. Limits on that are obtained \cite{review} from the endpoint of the tritium beta decay spectrum, from cosmology and from neutrinoless double beta decay ($0\nu \beta \beta$). From tritium we have an absolute upper limit of
2.2 eV (at 95\% C.L.) \cite{H3} on the antineutrino mass eigenvalues involved in beta decay, which, combined with the observed oscillation
frequencies under the assumption of three CPT-invariant light neutrinos, also amounts to an upper bound on the masses of
the other active neutrinos. Complementary information on the sum of neutrino masses is also provided by the galaxy power
spectrum combined with measurements of the cosmic  microwave background anisotropies. According to recent analyses of the most reliable data \cite{fo} one obtains 
$\sum_i \vert m_i\vert < 0.60\div 0.75$ eV (at 95\% C.L.), depending on the retained data and the cosmological model priors assumed. These numbers for the sum have to be divided by 3 in order to obtain a limit on the mass of each light neutrino.
The discovery of $0\nu \beta \beta$ decay would be very important because it would establish lepton number violation and
the Majorana nature of $\nu$'s, and provide direct information on the absolute
scale of neutrino masses.
The present limit from $0\nu \beta \beta$  (with large ambiguities from nuclear matrix elements) is about $\vert m_{ee}\vert < (0.3\div 0.8)$ eV \cite{rod} (see eq. (\ref{3nu1gen})). 

By now, after KamLAND, SNO and the upper limits on the absolute value of neutrino masses, not too much hierarchy in the spectrum of neutrinos is indicated by experiments \cite{Fogli}, \cite{Schwetz}: 
\bea
r = \Delta m_{sol}^2/\Delta m_{atm}^2=0.032\pm0.002 \sim 1/30.\label{r}
\eea
Thus, for a hierarchical spectrum, $m_2/m_3 \sim \sqrt{r} \sim 0.2$, which is comparable to $\lambda_C \sim 0.226$ (thoughout this article  $\lambda_C = \sin{\theta_C}$, with $\theta_C$ being the Cabibbo angle) or $\sqrt{m_{\mu}/m_{\tau}} \sim 0.24$. This suggests that the same hierarchy parameter (raised to powers with O(1) exponents) may apply for quark, charged lepton and neutrino mass matrices. In fact, $m_\mu/m_\tau \sim0.06 \sim \lambda_C^2$ and $m_e/m_\mu \sim 0.005\sim\lambda_C^{3-4}$).

For the near future the most important experimental challenges on neutrino oscillation experiments are more precise measurements of the absolute scale of neutrino mass (KATRIN, MARE), the accurate determination of $\theta_{13}$ (from MINOS, T2K and the reactor experiments DOUBLE CHOOZ, Daya Bay and RENO) and of the shift from maximal of $\theta_{23}$, the fixing of the sign of $\Delta m^2_{23}$ (normal or inverse hierarchy) (e.g. NO$\nu$A), the detection of CP violation in $\nu$ oscillations. Related to neutrino physics is the issue of the non conservation of the separate e, $\mu$ and $\tau$ lepton numbers. The recent new limit Br($\mu \rightarrow e \gamma) \lappeq 2.4^.10^{-12}$ obtained by the MEG experiment \cite{MEG} is largely satisfied in the SM but it imposes a strong constraint on SUSY-GUT models.

\section{Importance of Neutrinoless Double Beta Decay}

The detection of neutrino-less double beta decay \cite{Bar} would provide direct evidence of $L$ non conservation and of the Majorana nature of neutrinos. It would also offer a way to possibly disentangle the 3 cases of degenerate, normal or inverse hierachy neutrino spectrum.  The quantity which is bound by experiments on $0\nu \beta \beta$
is the 11 entry of the
$\nu$ mass matrix, which in general, from $m_{\nu}=U^* m_{diag} U^\dagger$, is given by :
\bea 
\vert m_{ee}\vert~=\vert(1-s^2_{13})~(m_1 c^2_{12}~+~m_2 s^2_{12})+m_3 e^{2 i\phi} s^2_{13}\vert
\label{3nu1gen}
\eea
where $m_{1,2}$ are complex masses (including Majorana phases) while $m_3$ can be taken as real and positive and $\phi$ is the $U_{PMNS}$ phase measurable from CP violation in oscillation experiments. Starting from this general formula it is simple to
derive the bounds for degenerate, inverse hierarchy or normal hierarchy mass patterns shown in Fig.1 \cite{fsv}.

\begin{figure}
\centering
\includegraphics[width=10cm]{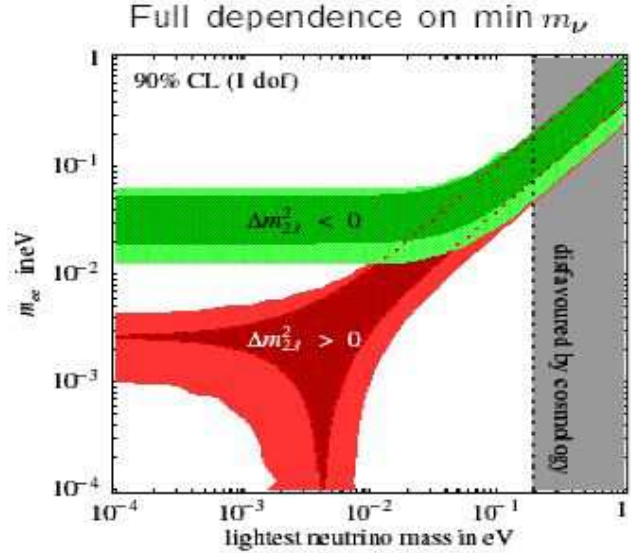}    
\caption[ ]{A plot \cite{fsv} of $m_{ee}$ in eV, the quantity measured in neutrino-less double beta decay, given in eq.(\ref{3nu1gen}), versus the lightest neutrino mass $m_1$, also in eV. The upper (lower) band is for inverse (normal) hierarchy.}
\end{figure}

In the next few years a new generation of experiments (CUORE, GERDA, ....) will reach a larger sensitivity on $0\nu \beta \beta$ by about an order of magnitude. Assuming the standard mechanism through mediation of a light massive Majorana neutrino, if these experiments will observe a signal this would indicate that the inverse hierarchy is realized, if not, then the normal hierarchy case remains a possibility. 

\section{Baryogenesis via Leptogenesis from Heavy $\nu^c$ Decay}

In the Universe we observe an apparent excess of baryons over antibaryons. It is appealing that one can explain the
observed baryon asymmetry by dynamical evolution (baryogenesis) starting from an initial state of the Universe with zero
baryon number.  For baryogenesis one needs the three famous Sakharov conditions: B violation, CP violation and no thermal
equilibrium. In the history of the Universe these necessary requirements have possibly occurred at different epochs. Note
however that the asymmetry generated during one such epoch could be erased in following epochs if not protected by some dynamical
reason. In principle these conditions could be fulfilled in the SM at the electroweak phase transition. In fact, when kT is of the order of a few TeV, B conservation is violated by
instantons (but B-L is conserved), CP symmetry is violated by the CKM phase and
sufficiently marked out-of- equilibrium conditions could be realized during the electroweak phase transition. So the
conditions for baryogenesis  at the weak scale in the SM superficially appear to be present. However, a more quantitative
analysis
\cite{tro} shows that baryogenesis is not possible in the SM because there is not enough CP violation and the phase
transition is not sufficiently strong first order, unless the Higgs mass is below a bound which by now is excluded by LEP. In SUSY extensions of the SM, in particular in the MSSM,
there are additional sources of CP violation and the bound on $m_H$ is modified but also this possibility has by now become at best marginal after the results from LEP2.

If baryogenesis at the weak scale is excluded by the data it can occur at or just below the GUT scale, after inflation.
But only that part with
$|{\rm B}-{\rm L}|>0$ would survive and not be erased at the weak scale by instanton effects. Thus baryogenesis at
$kT\sim 10^{10}-10^{15}~{\rm GeV}$ needs B-L violation and this is also needed to allow $m_\nu$ if neutrinos are Majorana particles.
The two effects could be related if baryogenesis arises from leptogenesis then converted into baryogenesis by instantons
\cite{buch}. The decays of heavy Majorana neutrinos (the heavy eigenstates of the see-saw mechanism) happen with violation of lepton number L, hence also of B-L and can well involve a sufficient amount of ¤CP violation. Recent results on neutrino masses are compatible with this elegant possibility. Thus the case
of baryogenesis through leptogenesis has been boosted by the recent results on neutrinos.

\section{Models of Neutrino Mixing}

\begin{figure}[]
\centering
$$\hspace{-4mm}
\includegraphics[width=10.0 cm]{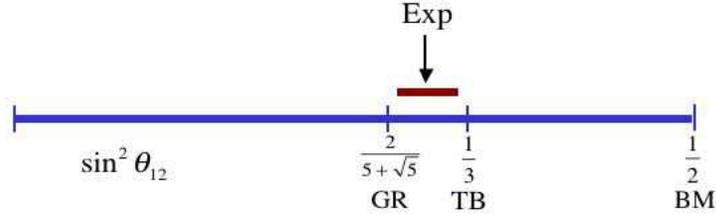}$$
\caption[]{The values of $\sin^2{\theta_{12}}$ for TB or GR or BM mixing are compared with the data }
\end{figure}

Neutrino mixing is important because it could in principle provide new clues for the understanding of the flavour problem. Even more so since neutrino mixing angles show a pattern that is completely different than that of quark mixing: for quarks all mixing angles are small, for neutrinos two angles are large (one is even compatible with the maximal value) and only the third one is small. We first consider the case of models based on discrete flavour groups that have received a lot of attention in recent years \cite{rmp}. There are a number of special mixing patterns that have been studied in this context. These mixing matrices all have $\sin^2{\theta_{23}}=1/2$, $\sin^2{\theta_{13}}=0$  and differ by the value of $\sin^2{\theta_{12}}$  (see Fig. 2). The corresponding mass matrices are 2-3 symmetric , i.e. $\mu-\tau$ symmetric (see, as examples, the early work in ref.\cite{fuk} and the recent paper ref.\cite{rompi}). The observed value of $\sin^2{\theta_{12}}$ \cite{Fogli}, \cite{Schwetz} the best measured mixing angle,  is very close, from below, to the so called Tri-Bimaximal (TB) value \cite{hps} which is $\sin^2{\theta_{12}}=1/3$. Alternatively it is also very close, from above, to the Golden Ratio (GR) value \cite{kaji}, \cite{ever}, \cite{fp} which is $\sin^2{\theta_{12}}=\frac{1}{\sqrt{5}\phi} = \frac{2}{5+\sqrt{5}}\sim 0.276$, where $\phi= (1+\sqrt{5})/2$ is the GR (for a different connection to the GR in this context, see \cite{rodgr}, \cite{adul}). Thus, a possibility is that one or the other of these coincidences is taken seriously and this leads to models where either TB or GR mixing is naturally predicted as a good first approximation. Here I will mainly refer to TB mixing which is the simplest and most studied first approximation to the data. 

The  TB mixing matrix (in a particular phase convention) is given by:
\begin{equation}
U_{TB}= \left(\matrix{
\dd\sqrt{\frac{2}{3}}&\dd\frac{1}{\sqrt 3}&0\cr
-\dd\frac{1}{\sqrt 6}&\dd\frac{1}{\sqrt 3}&-\dd\frac{1}{\sqrt 2}\cr
-\dd\frac{1}{\sqrt 6}&\dd\frac{1}{\sqrt 3}&\dd\frac{1}{\sqrt 2}}\right)~~~~~. 
\label{2}
\end{equation}

Note that the mixing angles are independent of mass ratios (while, for quark mixings, relations like $\lambda_C^2\sim m_d/m_s$ are typical). In the basis where charged lepton masses are 
diagonal, the effective neutrino mass matrix in the TB case is given by $m_{\nu}=U_{TB}\rm{diag}(m_1,m_2,m_3)U_{TB}^T$:
\begin{equation}
m_{\nu}=  \left[\frac{m_3}{2}M_3+\frac{m_2}{3}M_2+\frac{m_1}{6}M_1\right]~~~~~. 
\label{1k1}
\end{equation}
where:
\be
M_3=\left(\matrix{
0&0&0\cr
0&1&-1\cr
0&-1&1}\right),~~~~~
M_2=\left(\matrix{
1&1&1\cr
1&1&1\cr
1&1&1}\right),~~~~~
M_1=\left(\matrix{
4&-2&-2\cr
-2&1&1\cr
-2&1&1}\right).
\label{4k1}
\ee
The eigenvalues of $m_{\nu}$ are $m_1$, $m_2$, $m_3$ with eigenvectors $(-2,1,1)/\sqrt{6}$, $(1,1,1)/\sqrt{3}$ and $(0,1,-1)/\sqrt{2}$, respectively. The expression in eq.(\ref{1k1}) can be reproduced in models with sequential dominance or with form dominance, discussed by S. King and collaborators \cite{ski}. 

As we see the most general neutrino mass matrix corresponding to TB mixing, in the basis of diagonal charged leptons, is of the form:
\begin{equation}
m=\left(\matrix{
x&y&y\cr
y&x+v&y-v\cr
y&y-v&x+v}\right),
\label{gl21}
\end{equation}
This is a symmetric, 2-3 symmetric matrix with $a_{11}+a_{12}=a_{22}+a_{23}$.

We now discuss models that naturally produce TB mixing in first approximation. Discrete non-abelian groups naturally emerge as suitable flavour symmetries \cite{rmp}. In fact the TB mixing matrix immediately suggests rotations by fixed, discrete angles. In a series of papers, started by \cite{TBA4} (for a rather complete list of references see \cite{rmp}, some recent papers are in ref.\cite{recA4}) it has been pointed out that a broken flavour symmetry based on the discrete
group $A_4$ appears to provide a simplest realization of this specific mixing pattern in Leading Order (LO). We recall that $A_n$ is the group of even permutations of n objects (n!/2 elements). In the case of GR mixing the simplest choice is the group $A_5$ \cite{fp}. Other
solutions for TB mixing based on alternative discrete or  continuous flavour groups have also been considered \cite{rmp}, \cite{ishi} but the $A_4$ models have a very economical and attractive structure, e.g. in terms of group representations and of field content. 

We recall that $A_4$ can be generated by the two elements
$S$ and $T$ obeying the relations (a "presentation" of the group):
\be
S^2=(ST)^3=T^3=1~~~.
\label{$A_4$}
\ee
The 12 elements of $A_4$  are obtained as:
$1$, $S$, $T$, $ST$, $TS$, $T^2$, $ST^2$, $STS$, $TST$, $T^2S$, $TST^2$, $T^2ST$.
The inequivalent irreducible representations of $A_4$ are 1, 1', 1" and 3. Note that the squares of the dimensions of all these representations add up to 12, the dimension of $A_4$. It is immediate to see that one-dimensional unitary representations are
given by:
\be
\begin{array}{lll}
1&S=1&T=1\\
1'&S=1&T=e^{\dd i 4 \pi/3}\equiv\omega^2\\
1''&S=1&T=e^{\dd i 2\pi/3}\equiv\omega \label{s$A_4$}
\end{array}
\ee
The three-dimensional unitary representation, in a basis
where the element $T$ is diagonal, is given by:
\be
T=\left(
\begin{array}{ccc}
1&0&0\\
0&\omega^2&0\\
0&0&\omega
\end{array}
\right),~~~~~~~~~~~~~~~~
S=\frac{1}{3}
\left(
\begin{array}{ccc}
-1&2&2\cr
2&-1&2\cr
2&2&-1
\end{array}
\right)~~~.
\label{ST}
\ee

Note that the generic mass matrix for TB mixing in eq.(\ref{gl21}) can be specified as the most general matrix that is invariant under $\mu-\tau$ symmetry, implemented by the unitary matrix  $A_{\mu \tau}$:
\be
A_{\mu \tau}=\left(
\begin{array}{ccc}
1&0&0\\
0&0&1\\
0&1&0
\end{array}
\right)
\label{Amutau}
\ee
and under the $S$ transformation:
\bea
m=SmS,~~~~~m=A_{\mu \tau}mA_{\mu \tau}~~\label{inv}
\eea
where S is given in eq.(\ref{ST}).

The $m$ mass matrix of eq.(\ref{gl21}) is derived in the basis where charged leptons are diagonal. It is useful to consider the product $m^2=m_e^\dagger m_e$, where $m_e$ is the charged lepton mass matrix (defined as $\overline \psi_R m_e \psi_L$), because this product transforms as $m'^2=U_e^\dagger m^2 U_e$, with $U_e$ the unitary matrix that rotates the left-handed (LH) charged lepton fields. The most general diagonal $m^2$ is invariant under a diagonal phase matrix with 3 different phase factors:
\beq
m_e^\dagger m_e= T^\dagger m_e^\dagger m_e T
\label{Tdiag}
\eeq
and conversely a matrix $m_e^\dagger m_e$ satisfying the above requirement is diagonal. If $T^n=1$
the matrix $T$ generates a cyclic group $Z_n$.
The simplest case is $n=3$, which corresponds to $Z_3$ (but $n>3$ is equally possible) and to the $T$ matrix in eq.(\ref{ST}).

We can now see why $A_4$ works for TB mixing. It works because $S$ and $T$ are matrices of $A_4$ (in fact they satisfy eqs.(\ref{$A_4$})).  One could object that the matrix $A_{23}$ is not an element of $A_4$ 
(because the 2-3 exchange is an odd permutation). 
But it can be shown that in $A_4$ models the 2-3 symmetry is maintained by imposing that there are no flavons transforming as $1'$ or $1''$ that break $A_4$ with two different VEV's: in particular one can assume that there are no flavons in the model transforming as $1'$ or $1''$ \cite{AFextra}.

The group $A_4$ has two obvious subgroups: $G_S$, which is a reflection subgroup
generated by $S$ and $G_T$, which is the group generated by $T$, which is isomorphic to $Z_3$.
If the flavour symmetry associated to $A_4$ is broken by the VEV of a triplet
$\varphi=(\varphi_1,\varphi_2,\varphi_3)$ of scalar fields,
there are two interesting breaking pattern. The VEV
\be
\langle\varphi\rangle=(v_S,v_S,v_S)
\label{unotre}
\ee
breaks $A_4$ down to $G_S$, while
\be
\langle\varphi\rangle=(v_T,0,0)
\label{unozero}
\ee
breaks $A_4$ down to $G_T$. We have seen that $G_S$ and $G_T$ are the relevant low-energy symmetries
of the neutrino and the charged-lepton sectors, respectively. Indeed we have shown that the TB mass matrix is invariant under $G_S$
and, for charged leptons, a diagonal  $m_e^\dagger m_e$ is invariant under $G_T$. 
A crucial part of all serious A4 models is the dynamical generation of this alignment in a natural way.
In most of the models $A_4$ is accompanied by additional flavour symmetries, either discrete like $Z_N$ or continuous like U(1), which are necessary to eliminate unwanted couplings, to ensure the needed vacuum alignment (obtained from the minimization of the most general potential compatible with the assumed symmetries) and to reproduce the observed mass hierarchies. Explicit realizations of models for TB mixing based on $A_4$ can be found, for example, in \cite{AFextra,AFmodular,AFL,afh,altveram}. The possible origin of $A_4$ from a deeper level of the theory has been discussed in the context of extra dimensions and orbifolding \cite{AFL}, \cite{Adul} or as related to the fact that $A_4$ is a subgroup of the modular group \cite{AFmodular}, which plays a role in string theory. In passing, we note that we are mainly interested here in the possibility that the flavour symmetry is valid at the GUT scale and is broken at lower scales.

\section{Applying $A_4$ to lepton masses and mixings}

In the lepton sector a typical $A_4$ model works as follows \cite{AFmodular}.  One assigns
leptons to the four inequivalent
representations of $A_4$: LH lepton doublets $l$ transform
as a triplet $3$, while the RH charged leptons $e^c$,
$\mu^c$ and $\tau^c$ transform as $1$, $1''$ and $1'$, respectively. 
Here we consider a  see-saw realization, so we also introduce conjugate neutrino fields $\nu^c$ transforming as a triplet of $A_4$. The fact that LH lepton doublets $l$ and, in the see-saw case, also the RH neutrinos $\nu^c$, transform as triplets is crucial to realize the fixed ratios of mass matrix elements needed to obtain TB mixing. A drawback is that for the ratio $r$, defined in eq.(\ref{r}), one would expect $\sqrt{r} \approx \mathcal{O}(1)$ to be compared with the experimental value is $\sqrt{r} \approx 0.2$, which implies a moderate fine tuning. 

One adopts a supersymmetric (SUSY) context also to make contact 
with Grand Unification (flavor symmetries are supposed to act near the GUT scale). In fact, as well known, SUSY is important in GUT's for offering a solution to the hierarchy problem, for improving coupling unification and for making the theory compatible with bounds on proton decay. But, in models of lepton mixing, SUSY also helps for obtaining the vacuum alignment, because the SUSY constraints are very strong and limit the form of the superpotential very much. Thus SUSY is not necessary but it is a plausible and useful ingredient. The flavor symmetry is broken by two triplets
$\varphi_S$ and $\varphi_T$ (with the vacuum alignment in eqs.(\ref{unotre}, \ref{unozero})) and by one or more singlets $\xi$. All these fields are invariant under the SM gauge symmetry.
Two Higgs doublets $h_{u,d}$, invariant under $A_4$, are
also introduced. One can obtain  the observed hierarchy among $m_e$, $m_\mu$ and
$m_\tau$ by introducing an additional U(1)$_{FN}$ flavor symmetry \cite{Frogg} under
which only the  RH  lepton sector is charged (recently some models were proposed with a different VEV alignment such that the charged lepton hierarchies are obtained without introducing a $U(1)$ symmetry \cite{linfl,altveram}).
We recall that $U(1)_{FN}$ is a simplest flavor symmetry where particles in different generations are assigned (in general) different values of an Abelian charge.  Also Higgs fields may get a non zero charge. When the symmetry is spontaneously broken the entries of mass matrices are suppressed if there is a charge mismatch and more so if the corresponding mismatch is larger.
We assign FN-charges $0$, $q$ and $2q$ to $\tau^c$, $\mu^c$ and
$e^c$, respectively. There is some freedom in the choice of $q$.
Here we take $q=2$.
By assuming that a flavon $\theta$, carrying
a negative unit of FN charge, acquires a VEV 
$\langle \theta \rangle/\Lambda\equiv\lambda<1$, the Yukawa couplings
become field dependent quantities $y_{e,\mu,\tau}=y_{e,\mu,\tau}(\theta)$
and we have
\be
y_\tau\approx \mathcal{O}(1)~~~,~~~~~~~y_\mu\approx O(\lambda^2)~~~,
~~~~~~~y_e\approx O(\lambda^{4})~~~.
\ee
Had we chosen $q=1$, we would have needed  $\langle \theta \rangle/\Lambda$ of order $\lambda^2$, to reproduce the above result.
The superpotential term for lepton masses, $w_l$ is given by:
\be
w_l=y_e e^c (\varphi_T l)+y_\mu \mu^c (\varphi_T l)'+
y_\tau \tau^c (\varphi_T l)''+ y (\nu^c l)+
(x_A\xi+\tilde{x}_A\tilde{\xi}) (\nu^c\nu^c)+x_B (\varphi_S \nu^c\nu^c)+...
\label{wlss}
\ee
with dots denoting higher   
dimensional operators that lead to corrections to the LO   
approximation. In our notation, the product of 2 triplets  $(3 3)$ transforms as $1$, 
$(3 3)'$ transforms as $1'$ and $(3 3)''$ transforms as $1''$. 
To keep our formulae compact, we omit to write the Higgs and flavon  fields
$h_{u,d}$, $\theta$ and the cut-off scale $\Lambda$. For instance 
$y_e e^c (\varphi_T l)$ stands for $y_e e^c (\varphi_T l) h_d \theta^4/\Lambda^5$. The parameters of the superpotential $w_l$ are complex, in particular those responsible for the
heavy neutrino Majorana masses, $x_{A,B}$. Some terms allowed by the $A_4$ symmetry, such as the terms 
obtained by the exchange $\varphi_T\leftrightarrow \varphi_S$, 
(or the term $(\nu^c\nu^c)$) are missing in $w_l$. 
Their absence is crucial and, in each version of $A_4$ models, is
motivated by additional symmetries. 

As for the neutrino spectrum both normal and inverted hierarchies 
can be realized. It is interesting that $A_4$ models with the see-saw mechanism typically lead to a light neutrino spectrum which satisfies the sum rule (among complex masses):
\be
\frac{1}{m_3}=\frac{1}{m_1}-\frac{2}{m_2}~~~.\\
\label{sumr}
\ee
A detailed discussion of a spectrum of this type can be found in refs. \cite{AFmodular,altveram}.
The above sum rule gives rise to bounds on the lightest neutrino mass.
As a consequence, for example, the possible values of $|m_{ee}|$ are restricted. For normal hierarchy we have
\be
|m_{ee}|\approx \dd\frac{4}{3\sqrt{3}} \Delta m^2_{sun} \approx 0.007~{\rm eV}~~~.
\label{meeno}
\ee
while for inverted hierarchy
\be
|m_{ee}|\ge  \dd\sqrt{\frac{\Delta m^2_{atm}}{8}} \approx 0.017~{\rm eV}~~~.
\label{meeio}
\ee
In a completely general framework, without the restrictions imposed by the flavor symmetry,
$|m_{ee}|$ could vanish in the case of normal hierarchy. In this model $|m_{ee}|$ is always
different from zero, though its value for normal hierarchy is probably too small to be detected
in the next generation of $0\nu\beta\beta$ experiments.

In the leading approximation $A_4$ models lead to exact TB mixing.  In these models TB mixing is implied by the symmetry at the leading order approximation which is corrected by non leading effects. Given the set of flavour symmetries and having specified the field content, the non leading corrections to TB mixing, arising from higher dimensional effective operators, can be evaluated in a well defined expansion. In the absence of specific dynamical tricks, in a generic model, all the three mixing angles receive corrections of the same order of magnitude. Since the experimentally allowed departures of $\theta_{12}$ from the TB value, $\sin^2{\theta_{12}}=1/3$, are small, numerically not larger than $\mathcal{O}(\lambda_C^2)$, it follows that both $\theta_{13}$ and the deviation of $\theta_{23}$ from the maximal value are expected   to also be typically of the same general size. The central values $\sin{\theta_{13}} \sim 0.16~-~0.11$ that can be derived from the experimental results in the two columns of Table(\ref{tab:data}), respectively, are in between  $\mathcal{O}(\lambda_C^2) \sim \mathcal{O}(0.05)$ and $\mathcal{O}(\lambda_C) \sim \mathcal{O}(0.23)$. Thus models based on TB or GR mixing are still viable with preference for the lower side of the experimental range for $\theta_{13}$. It is also to be noted that one can introduce some additional theoretical input to enhance the value of $\theta_{13}$. In the case of $A_4$, examples are provided by the model of ref.\cite{linx}, formulated before the T2K and MINOS results were known and the modified $A_4$ model of ref.\cite{A4mod} (see also \cite{merva}).

\section{$A_4$, quarks and GUT's}

Much attention  has been devoted to the question whether models with TB mixing in the neutrino sector can be  suitably extended to also successfully describe the observed pattern of quark mixings and masses and whether this more complete framework can be made compatible with (supersymmetric) SU(5) or SO(10) Grand Unification. 

The simplest attempts of directly extending models based on $A_4$ to quarks have not been  satisfactory.
At first sight the most appealing
possibility is to adopt for quarks the same classification scheme under $A_4$ that one has
used for leptons (see, for example, ref.\cite{AFmodular}). Thus one tentatively assumes that LH quark doublets $Q$ transform
as a triplet $3$, while the  antiquarks $(u^c,d^c)$,
$(c^c,s^c)$ and $(t^c,b^c)$ transform as $1$, $1''$ and $1'$, respectively. This leads to $V_u=V_d$ and to the identity matrix for $V_{CKM}=V_u^\dagger V_d$ in the lowest approximation. This at first appears as very promising: a LO approximation where neutrino mixing is TB and $V_{CKM}=1$ is a very good starting point. But there are some problems. First, the corrections 
to $V_{CKM}=1$ turn out to be strongly constrained by the leptonic sector, because lepton mixing angles are very close to the TB values, and, in the simplest models, this constraint leads to a too small $V_{us}$
(i.e. the Cabibbo angle is rather large in comparison to the allowed shifts from the TB mixing angles). Also in these models, the quark classification which leads to $V_{CKM}=1$ is not compatible with $A_4$ commuting with SU(5). 
An additional consequence of the above assignment is that the top quark mass arises from a non-renormalizable dimension-5 operator. In that case, to reproduce the top mass, we need 
to compensate the cutoff suppression by some extra dynamical mechanism. Alternatively, we have to introduce a separate symmetry breaking parameter for the quark sector, sufficiently close to the cutoff
scale.

Due to this, larger discrete groups have been considered for the description of quarks.
A particularly appealing set of models is based on the discrete group $T'$, the double covering group of $A_4$ \cite{Tpr}, \cite{Feruglio:2007}, \cite{Chen:2007}. The 
representations of $T'$ are those of $A_4$ plus three independent doublets 2, $2'$ and $2''$. The doublets are interesting for the classification of the first two generations of quarks \cite{su2}. For example, in ref.\cite{Feruglio:2007} a viable description was obtained, i.e. in the leptonic sector the predictions of the $A_4$ model are maintained, while the $T'$ symmetry plays an essential role for reproducing the pattern of quark mixing. But, again, the classification adopted in this model is not compatible with Grand Unification.

As a result, the group $A_4$ was considered by many authors to be too
limited to also describe quarks and to lead to a grand unified
description. But it has been shown \cite{afh} that this negative attitude
is not justified and that it is actually possible to construct a
viable model based on $A_4$ which leads to a grand
unified theory (GUT) of quarks and leptons with TB mixing
for leptons and with quark (and charged lepton) masses and mixings compatible with experiment. At the same time this model offers an example of an
extra dimensional SU(5) GUT in which a description of all fermion masses
and mixings is accomplished.  The
formulation of SU(5) in extra dimensions has the usual advantages of
avoiding large Higgs representations to break SU(5) and of solving the
doublet-triplet splitting problem.  The choice of the transformation properties of the two
Higgses $H_5$ and $H_{\overline{5}}$ has a special role in this model. They are chosen to transform 
as two different $A_4$ singlets
$1$ and $1'$. As a consequence, mass terms for the Higgs colour
triplets are  not directly allowed and their masses are
introduced by orbifolding, \`{a} la Kawamura \cite{Kawamura:2001}.  In this model, proton
decay is dominated by gauge vector boson exchange giving rise to
dimension-6 operators, while the usual contribution of dimension-5 operators is forbidden by the selection rules of the model. Given the large $M_{GUT}$ scale of SUSY models and the relatively huge theoretical uncertainties, the decay rate is within the present experimental limits.
A see-saw realization
in terms of an $A_4$ triplet of RH neutrinos $\nu^c$ ensures the
correct ratio of light neutrino masses with respect to the GUT
scale. In this model extra dimensional effects directly
contribute to determine the flavour pattern, in that the two lightest
tenplets $T_1$ and $T_2$ are in the bulk (with a doubling $T_i$ and
$T'_i$, $i=1,2$ to ensure the correct zero mode spectrum), whereas the
pentaplets $F$ and $T_3$ are on the brane. The hierarchy of quark and
charged lepton masses and of quark mixings is determined by a
combination of extra dimensional suppression factors and of $U(1)_{FN}$ charges, both of which only apply to the first two
generations, while the neutrino mixing angles
derive from $A_4$ in the usual way. If the extra dimensional suppression factors and the $U(1)_{FN}$ charges are switched off, only the third generation masses of quarks and charged leptons survive. Thus the charged fermion mass matrices are nearly empty in this limit (not much of $A_4$ effects remain) and the quark mixing angles are determined by the small corrections induced by the above effects. The model is natural, since most of the
small parameters in the observed pattern of masses and mixings as well
as the necessary vacuum alignment are  justified by the symmetries of
the model. However, in this case, like in all models based on $U(1)_{FN}$, the number of $\mathcal{O}(1)$ parameters is larger than the number of measurable quantities, so that in the quark sector the model can only account for the orders of magnitude (measured in terms of powers of an expansion parameter) and not for the exact values of mass ratios and mixing angles. A moderate fine tuning is only needed to enhance the Cabibbo mixing angle between the first two generations, which would generically be of $\mathcal{O}(\lambda_C^2)$. 

The problem of constructing GUT models based on  $SU(5)\otimes G_f$ or $SO(10)\otimes G_f$ with approximate TB mixing in the leptonic sector has also been considered by many authors. Examples are: for $G_f=A_4$ ref.\cite{a4gut}, for $T'$ ref.\cite{Chen:2007}, for $S_4$ ref.\cite{ishim}. 
As for the models based on $SO(10)\otimes G_f$  recent examples were discussed with $G_f=S_4$ \cite{Dutta:2009} and $G_f=PSL_2(7)$ \cite{King:2009a}. Clearly the case of $SO(10)$ is even more difficult than that of $SU(5)$ because the neutrino sector is tightly related to that of quarks and charged leptons as all belong to the 16 of $SO(10)$. For a discussion of $SO(10)\otimes A_4$ models, see \cite{Bazzocchi:2008b}. More in general see refs.\cite{alro1}.
In our opinion most of the models are incomplete (for example, the crucial issue of VEV alignment is not really treated in depth as it should) and/or involve a number of unjustified steps and ad-hoc fine tuning of parameters. 

While $A_4$ is the minimal flavor group leading to TB mixing, alternative flavor groups have been studied in the literature and can lead to interesting variants with some specific features.
Actually, in ref.\cite{lam}, the claim was made that, in order to obtain the TB mixing "without fine tuning", the finite group must be $S_4$ or a larger group containing $S_4$. For us this claim is not well grounded being based on an abstract mathematical criterium for a natural model (see also \cite{gri}). For us a physical field theory model is natural if the interesting results are obtained from the most general lagrangian compatible with the stated symmetry and the specified representation content for the flavons. For example, we obtain from $A_4$ (which is a subgroup of $S_4$) a natural (in our sense) model for the TB mixing by simply not including symmetry breaking flavons transforming like the $1'$ and the $1''$ representations of $A_4$. This limitation on the transformation properties of the flavons is not allowed by the rules specified in ref.\cite{lam} which demand that the symmetry breaking is induced by all possible kinds of flavons (note that, according to this criterium, the SM of electroweak interactions would not be natural because only Higgs doublets are introduced!). Rather, for naturalness we also require that additional physical properties like the VEV alignment or the hierarchy of charged lepton masses also follow from the assumed symmetry and are not obtained by fine tuning parameters: for this actually $A_4$ can be more effective than $S_4$ because it possesses three different singlet representations 1, $1'$ and $1''$.  

Models of neutrino mixing based on $S_4$ have in fact been studied \cite{s4}. The group of the permutations of 4 objects $S_4$ has 24 elements and 5 equivalence classes that correspond to 5 inequivalent irreducible representations, two singlets, one doublet, two triplets: $1_1$, $1_2$, $2$, $3_1$ and $3_2$. Note that the squares of the dimensions of all these representations add up to 24. 

\section{Bimaximal Mixing and S4}

The new results showing that probably $\theta_{13}$ is not far from its former upper bound could alternatively be interpreted as an indication that the agreement with the TB or GR mixing is accidental. Then a scheme where instead the Bimaximal (BM) mixing is the correct first approximation modified by terms of $\mathcal{O}(\lambda_C)$ could be relevant. In BM mixing $\theta_{12}$ and $\theta_{23}$ are both maximal while $\theta_{13}=0$ (see. Fig. 2). This is in line with the well known empirical observation that $\theta_{12}+\theta_C\sim \pi/4$, a relation known as quark-lepton complementarity \cite{compl}, or similarly $\theta_{12}+\sqrt{m_\mu/m_\tau} \sim \pi/4$. No compelling model leading, without parameter fixing, to the exact complementarity relation has been produced so far. Probably the exact complementarity relation becomes more plausible if replaced with $\theta_{12}+\mathcal{O}(\theta_C)\sim \pi/4$ or $\theta_{12}+\mathcal{O}(m_\mu/m_\tau)\sim \pi/4$ (which we could call "weak" complementarity). One can think of models where, because of a suitable symmetry,  BM mixing holds in the neutrino sector at leading order and the necessary, rather large, corrective terms for $\theta_{12}$ arise from the diagonalization of charged lepton masses \cite{compl}. These terms of order $\mathcal{O}(\lambda_C)$ from the charged lepton sector would then generically also affect $\theta_{13}$ and the resulting value could well be compatible with the present experimental values of $\theta_{13}$. A word of caution must be kept in mind: in the presence of these relatively large off diagonal terms in the charged lepton diagonalizing matrix one must arrange that not too large contributions to the decays $\mu \rightarrow e \gamma$ or $\tau \rightarrow \mu \gamma$ are generated \cite{rmp}.

The BM mixing matrix is given by:
\begin{equation}
U_{BM}= \left(\matrix{
\dd\frac{1}{\sqrt 2}&\dd-\frac{1}{\sqrt 2}&0\cr
\dd\frac{1}{2}&\dd\frac{1}{2}&-\dd\frac{1}{\sqrt 2}\cr
\dd\frac{1}{2}&\dd\frac{1}{2}&\dd\frac{1}{\sqrt 2}}\right)\;.
\label{21}
\end{equation}

Along this line of thought, we have used the expertise acquired with non Abelian finite flavour groups to construct a model \cite{S4us} based on the permutation group $S_4$ which naturally leads to the BM mixing at leading order. We have adopted a supersymmetric formulation of the model in 4 space-time dimensions. The complete flavour group is $S_4\times Z_4 \times U(1)_{FN}$. In leading order, the charged leptons are diagonal and hierarchical and the light neutrino mass matrix, after see-saw, leads to the exact BM mixing. The model is built in such a way that the dominant corrections to the BM mixing, from higher dimensional operators in the superpotential,  only arise from the charged lepton sector at next-to-the-leading-order and naturally inherit $\lambda_C$ (which fixes the charged lepton mass hierarchies) as the relevant expansion parameter. As a result the mixing angles deviate from the BM values by terms of  $\mathcal{O}(\lambda_C)$ (at most), and weak complementarity holds. A crucial feature of the model is that only $\theta_{12}$ and $\theta_{13}$ are corrected by terms of $\mathcal{O}(\lambda_C)$ while $\theta_{23}$ is unchanged at this order (which is essential for a better agreement of the model with the present data). Recently the model was extended to include quarks in a $SU(5)$ Grand Unified version \cite{melo} or in a Pati-Salam framework \cite{meba}. An $SO(10)$ model is discussed in ref.\cite{pat}. 

\section{Anarchy versus Symmetry} 

We now briefly turn to models that do not take seriously any of the coincidences described above (i.e. the proximity of the data to the TB or GR patterns or the quark-lepton complementarity: these patterns cannot all be true and it is possible that none of them is true) and are therefore based on a less restrictive flavour symmetry. It is clear that the T2K hint that $\theta_{13}$ may be large is great news for the most extreme position of this type, which is "anarchy" \cite{hmw}: no symmetry at all in the lepton sector, only chance. This view predicts generic mixing angles, so the largest angle, $\theta_{23}$, should somewhat deviate from maximal and the smallest angle, $\theta_{13}$, should be as large as possible within the experimental bounds. Anarchy can be formulated in a $SU(5) \bigotimes U(1)$ context by taking different Froggatt-Nielsen charges \cite{Frogg} only for the $SU(5)$ tenplets (for example 10: (3,2,0) where 3 is the charge of the first generation, 2 of the second, zero of the third) while no charge differences appear in the $\bar 5$: $\bar 5$: (0,0,0). This assignment is in agreement with the empirical fact that the mass hierarchies are more pronounced for up quarks in comparison with down quarks and charged leptons. In a non see-saw model, with neutrino masses dominated by the contribution of the dimension-5 Weinberg operator in eq.(\ref{O5}), the $\bar 5$ vanishing charges directly lead to random neutrino mass and mixing matrices. In anarchical see-saw models also the charges of the SU(5) singlet RH neutrinos must be undifferentiated among the 3 generations: $1$: (0,0,0). Anarchy can be mitigated by assuming that it  only holds in the 2-3 sector: e.g $\bar 5$: (2,0,0) with the advantage that the first generation masses and the angle $\theta_{13}$ are then naturally small (see ref.\cite{desy} for a recent discussion of this model). In models with see-saw one can alternatively play with the charges for the RH SU(5) singlet neutrinos. If, for example, we take 1: (1, -1, 0), together with $\bar 5$: (2,0,0),  it is possible to get a normal hierarchy model with $\theta_{13}$ small and also with $r = \Delta m^2_{solar}/\Delta m^2_{atm}$ naturally small (see, for example, ref.\cite{afm}). In summary anarchy and its variants, all based on chance, offer a rather economical class of models that are among those that are compatible with the recent $\theta_{13}$ results, with preference with the upper side of the experimentally allowed range.

\section{Conclusion}

In the last decade we have learnt a lot about neutrino masses and mixings.  A list of important conclusions have been reached. Neutrinos are not all massless but their masses are very small. Probably masses are small because neutrinos are Majorana particles
with masses inversely proportional to the large scale M of lepton number violation. It is quite remarkable that M is empirically not far from $M_{GUT}$, so that
neutrino masses fit well in the SUSY GUT picture. Also out of equilibrium decays with CP and L violation of heavy RH neutrinos can produce a B-L asymmetry, then converted near the weak scale by instantons into an amount of B asymmetry compatible with observations (baryogenesis via leptogenesis) \cite{buch}.  It has been established that most probably active neutrinos are not a significant component of dark matter in the Universe. We have also understood there there is no contradiction between large neutrino mixings and small quark mixings, even in the context of GUTs.  

This is a very impressive list of achievements. Coming to a detailed analysis of neutrino masses and mixings a long collection of models have been formulated over the years. 
With continuous improvements of the data and more precise values of the mixing angles most of the models have been discarded by experiment. Still the surviving models span a wide range going from a maximum of symmetry, with discrete non-abelian flavour groups, to the opposite extreme of anarchy. By now, besides the detailed knowledge of the entries of the $V_{CKM}$ matrix we also have a reasonable determination of the neutrino mixing matrix $U_{PMNS}$. The data appear to suggest some special patterns (recall Fig. 2) like TB or GR or BM mixing to be valid in some leading approximation, corrected by small non leading terms. If one takes these "coincidences" seriously, then non-abelian discrete flavour groups emerge as the main road to an understanding of this mixing pattern. Indeed the entries of e.g. TB mixing matrix are clearly suggestive of "rotations" by simple, very specific angles. It is remarkable that neutrino and  quark mixings have such a different qualitative pattern.  An obvious question is whether some additional indication for discrete flavour groups can be obtained by considering the extension of the models to the quark sector, perhaps in a Grand Unified context. The answer appears to be that, while the quark masses and mixings can indeed be reproduced in models where TB or BM mixing is realized in the leptonic sector through the action of discrete groups, there are no specific additional hints in favour of discrete groups that come from the quark sector. Further important input could come from $\mu \rightarrow e \gamma$ and in general from lepton flavour violating processes, from $b \rightarrow s \gamma$ and from LHC physics. In fact, new physics at the weak scale could have important feedback on the physics of neutrino masses and mixing.

In conclusion, one could have imagined that neutrinos would bring a decisive boost towards the formulation of a comprehensive understanding of fermion masses and mixings. In reality it is frustrating that no real illumination was sparked on the problem of flavour. We can reproduce in many different ways the observations, in a wide range that goes from anarchy to discrete flavour symmetries) but we have not yet been able to single out a unique and convincing baseline for the understanding of fermion masses and mixings. In spite of many interesting ideas and the formulation of many elegant models the mysteries of the flavour structure of the three generations of fermions have not been much unveiled. 

I imagine that by the next edition of this by now classic School, we will know the value of $\theta_{13}$ with a better accuracy, from the continuation of T2K, MINOS and DOUBLE CHOOZ and from the start of Daya Bay and RENO. Some existing models will be eliminated and the surviving ones will be updated to become more quantitative in order to cope with a precisely known mixing matrix. A definitely non vanishing $\theta_{13}$ value will encourage the planning of long baseline experiments for the detection of CP violation in neutrino oscillations. Along the way the important issue of the existence of sterile neutrinos must be clarified. The on going or in preparation experiments on the absolute value of neutrino masses, on $0\nu \beta \beta$, on $\mu \rightarrow e \gamma$, on the search for dark matter etc can also lead to extremely important developments in the near future. So this field is very promising and there are all reasons to expect an exciting time ahead of us.

\section{Acknowledgments}

I warmly thank Prof. Antonino Zichichi for his kind invitation and for the impressive hospitality received in Erice, as usual. I also thank the Staff of the E. Majorana School, in particular Mrs. Fiorella Ruggiu for their assistance in organizing and managing my stay at the School. I am indebted with Luca Merlo for his critical reading of a preliminary version of this work. I am also glad to acknowledge interesting discussions on this subject with him and with Ferruccio Feruglio and Davide Meloni. Supported in part by PRIN 2008 and by the EU network LHCPHENONET.

\end{document}